\documentclass[a4paper]{jpconf}
\usepackage{graphicx}
\begin{document}
\title{Bell gems naturally split dynamics information from $SU(2^{2d}) \rightarrow U(1)^{2^{2d-1}-1} \times SU(2)^{2^{2d-1}}$}

\author{Francisco Delgado}

\address{Departamento de F\'isica y Matem\'aticas, Escuela de Ingenier\'ia y Ciencias, Tecnol\'ogico de Monterrey, Campus Estado de M\'exico, Atizap\'an, Estado de M\'exico, CP. 52926, M\'exico.}

\ead{fdelgado@itesm.mx}

\begin{abstract}
Quantum Computation and Quantum Information are continuously growing research areas which are based on nature and resources of quantum mechanics, as superposition and entanglement. In its gate array version, the use of convenient and appropriate gates is essential. But while those proposed gates adopt convenient forms for computational algorithms, in the practice, their design depends on specific quantum systems and stuff being used. Gates design is restricted to properties and limitations of interactions and physical elements being involved, where Quantum Control plays a deep role. Quantum complexity of multipartite systems and their interactions requires a tight control to manipulate their quantum states, either local and non-local ones, but still a reducibility procedure should be addressed. This work shows how a general $2d$-partite two level spin system in $SU(2d)$ could be decomposed in $2^{n-1}$ subsystems on $SU(2)$, letting establish control operations. In particular, it is shown that Bell gems basis is a set of natural states on which decomposition happen naturally under some interaction restrictions. Thus, alternating the direction of local interaction terms in the Hamiltonian, this procedure states a universal exchange semantics on those basis. The structure developed could be understood as a splitting of the $2d$ information channels into $2^{2d-1}$ pairs of $2$ level information subsystems.
\end{abstract}

\section{Introduction}
Quantum manipulation applied to information processing or information storage is actually a continuously growing research area searching for practical, easy and optimal solutions in evolution of quantum systems. For spin based resources, single two level systems on $SU(2)$ have well known exact and optimal control solutions in terms of energy or time \cite{dales1,boscain1}. Despite, bigger spin systems in general have a complex behavior still not completely known contrasting with those $SU(2)$ solutions. Spin interactions have been analyzed in terms of transference and control of entanglement in bipartite qubits \cite{terzis1}, chains and lattices \cite{niem1,pfeuty1,stelma1,novo1}. On these arrangements, several approaches have extended research on more complex systems depending on external parameters (temperature, strength of external fields, geometry, etc.) \cite{arnesen1, wang1, khaneja1}. 

Anysotropic Ising model for bipartite systems in $SU(4)$ \cite{delgadoA} lets a $U(1) \times SU(2)^2$ block decomposition when their evolution is written in a non-local basis instead of traditional computational basis. It means $\mathcal{H}^2$ becomes a direct sum of two subspaces, each one generated by a pair of Bell states. While, $U$ becomes in the semi-direct product $U(1) \times SU(2)^2$. In these terms, control can be reduced to two $SU(2)$ control problems in each block and exact solutions can be found \cite{delgadoB}. Blocks can be configured as a function of the direction of external driven interactions being included. Such scheme lets control transformations between pairs of Bell states on demand, therefore on the complete state in spite of the possibility to reconfigure those pairs. Thus, the procedure sets a control method to manipulate quantum information on matter based on Bell states as computational grammar instead of traditional computational basis, letting the transformation between any pair of elements of this basis under well known control procedures for $SU(2)$. Thus, reduction or decomposition schemes from large systems in terms of simpler problems based on isolated two level subsystems could to state easier and universal (but not globally optimal in general) control procedures to manipulate them. 

\section{Generalization of SU(2) decomposition}
Complexity of multiqubit systems grows unpredictably with their size in terms of entanglement properties, usefulness and control. In particular, their manipulation do not exhibit scalable rules departing from their smallest systems on which they are constituted. Recently, a series of works considering driven magnetic fields together with bipartite Ising interaction show the evolution matrix expressed in a non-local basis splits the dynamics on two weak related information subsystems. The splitting could be selected as function of field direction. In this sense, current work shows the generalization of that decomposition scheme for an extended kind of $n$-partite two level spin systems when is based on non-local generalized Bell states, the Bell gems basis \cite{jaeg1,sych1}. Structure developed could be understood as a splitting of the $n$ quantum information channels into $2^{n-1}$ information channels weakly related, in the form of two level subsystems. 

Problem stated in this work is established for a general Hamiltonian for $n$ coupled two level systems on $U(2^n)$. If it is written as a linear combination of tensor products of Pauli matrices:

\begin{eqnarray}\label{hamiltonian}
{H} = \sum_{\{i_k\}}{h_{\{i_k\}}} \bigotimes_{k=1}^n \sigma_{i_k} = \sum_{\mathcal{I}=0}^{4^n-1}{h_{\mathcal{I}^n_4}} \bigotimes_{k=1}^n \sigma_{\mathcal{I}^n_{4,k}}
\end{eqnarray}

\noindent where $\{i_k\}=\{i_1,i_2,...,i_n\}$, $i_k=0,1,2,3$ and ${h_{\{i_k\}}}$, a set of time dependent real functions in general. Sometimes, as in the second expression in (\ref{hamiltonian}), $\{i_k\}$ will be represented as a number $\mathcal{I} \in \{0,1,...,4^n-1\}$ when it is expressed in base-4 with $n$ digits, $\mathcal{I}^n_4$. $\mathcal{I}^n_{4,k}=i_k$ is its $k^{\rm{th}}$ term in that base. Additionally, $k=1,2,...,n$ and $\sigma_{i}$ for $i=0,1,2,3$ are respectively the unitary matrix and traditional Pauli matrices assumed being expressed for the computational basis $\left| 0 \right>, \left| 1 \right> \in \mathcal{H}^2$ of each part. Then, it represents a generalized model for spin and polarization systems obeying the Schr\"odinger equation for its associated evolution operator $U$. Without loss of generality, the identity term $h_{0...0}\sigma_0 \otimes \sigma_0 ... \sigma_0$ can be dropped because it only contributes with a global phase. Thus, Hamiltonian is traceless, remaining evolution operator belongs to $SU(2^n)$ and it can be written as a linear combination of eigenvectors projectors $\left| b_j \right>\left< b_j \right|$:

\begin{eqnarray}\label{H2}
H &=& \sum_{j=1}^{2^n} E_j \left| b_j \right>\left< b_j \right|
\end{eqnarray}

\noindent Clearly, these eigenstates become invariant under time evolution. $SU(2)$ decomposition, as it was obtained in \cite{delgadoA} can be induced in general by pairing eigenvalues under an arbitrary rule and then considering $2^n$ orthogonal states $\{ \left| \alpha_i \right> \}$ obtained by pairs as rotations of each defined pair of eigenvalues:

\begin{eqnarray}\label{alphas}
\left| b_{2i-1} \right> = A_i \left| \alpha_{j(i)} \right> + B_i \left| \alpha_{k(i)} \right> & \qquad \rightarrow \qquad \left| \alpha_{j(i)} \right> = A^*_i \left| b_{2i-1} \right> - B_i \left| b_{2i} \right> \nonumber \\
\left| b_{2i} \right> = -B^*_i \left| \alpha_{j(i)} \right> + A^*_i \left| \alpha_{k(i)} \right> & \qquad \rightarrow \qquad \left| \alpha_{k(i)} \right> = B^*_i \left| b_{2i-1} \right> + A_i \left| b_{2i} \right>
\end{eqnarray}

These transformations directly split the whole Hilbert space ${\mathcal H}^{2n}$ in a direct sum of subspaces generated by each pair of states and alternatively by the corresponding pair of eigenstates. There are lots of possibilities for last states in terms of eigenstates and pairings. For them, separability or entanglement properties are not necessarily assured as in \cite{delgadoA}, which corresponds only to a particular interaction. This new basis transforms the diagonal structure of Hamiltonian on the basis $\{ \left| b_{i} \right> \}$ into a $2 \times 2$ block structure on the basis $\{ \left| \alpha_j \right> \}$:

\begin{eqnarray}\label{sh2}
H &=& \sum_{i=1}^{2^{n-1}} {{\bf S}_H}_i = \left(
\begin{array}{c|c|c|c}
{{\bf S}_H}_1 & \bf{0} & ... & \bf{0}      \\
\hline
\bf{0} & {{\bf S}_H}_2  &  ... & \bf{0}     \\
\hline
\vdots     & \vdots        &  \ddots & \vdots         \\
\hline
\bf{0} & \bf{0}    & ...  & {{\bf S}_H}_{2^{n-1}} 
\end{array}
\right)
\end{eqnarray}

\noindent With this, each block is a $2 \times 2$ "localized" Hamiltonian which can be expressed on a Pauli-like basis in $U(2)=U(1) \times SU(2)$: ${\bf I}_i,{\bf X}_i,{\bf Y}_i$ and ${\bf Z}_i$:

\begin{eqnarray}\label{sh}
{{\bf S}_H}_i &=& \Delta_i^+ {\bf I}_i -2 {r_{A_i}} {r_{B_i}} \Delta^-_i \cos \Gamma_i {\bf X}_i + 
2 {r_{A_i}} {r_{B_i}}  \Delta^-_i \sin \Gamma_i {\bf Y}_i - ({r_{A_i}}^2-{r_{B_i}}^2) \Delta^-_i {\bf Z}_i \\
&\equiv& \Delta_i^+ {\bf I}_i + {{\bf S}_H}_i^0 \nonumber
\end{eqnarray}

\noindent where:

\begin{eqnarray}
&A_i = {r_{A_i}} e^{i \gamma_{A_i}} , B_i = {r_{B_i}} e^{i \gamma_{B_i}} \nonumber \\
&\Delta^{\pm}_i = \frac{1}{2 \hbar} (E_{2i} \pm E_{2i-1}) \nonumber \\
&\Gamma_i=\gamma_{A_i} - \gamma_{B_i}
\end{eqnarray}

In addition, this structure simplifies and translates quantum error correction procedures into specific flip errors. It means, $H$ can be written as a sum of Hamiltonian operators $H_i$ on different two level subspaces $\mathcal{H}_i^{2}$. This block structure is inherited to the evolution matrix via the $\tau$-time ordered integral:

\begin{eqnarray}
{{\bf S}_U}_i &=& \tau \{ e^{-\frac{i}{\hbar} \int_{0}^{t} {{\bf S}_H}_i} dt' \} = e^{-i \Delta_i^+ t} \tau \{ e^{-\frac{i}{\hbar} \int_{0}^{t} {{\bf S}_H}_i^0} dt' \}
\end{eqnarray}

\noindent assuring the possibility to apply $SU(2)$ quantum control optimal schemes:

\begin{eqnarray}
U &=& \sum_{i=1}^{2^{n-1}} {{\bf S}_U}_i = \left(
\begin{array}{c|c|c|c}
{{\bf S}_U}_1 & \bf{0} & ... & \bf{0}      \\
\hline
\bf{0} & {{\bf S}_U}_2  &  ... & \bf{0}     \\
\hline
\vdots     & \vdots        &  \ddots & \vdots         \\
\hline
\bf{0} & \bf{0}    & ...  & {{\bf S}_U}_{2^{n-1}} 
\end{array}
\right)
\end{eqnarray}

It implies that $U \in U(1)^{2^{n-1}-1} \times SU(2)^{2^{n-1}}$ (because one arbitrary factor phase, $e^{-i \Delta_i^+ t}$, in some block depends on remaining phase factors in spite of $U \in SU(2^n)$). Informally, we will call to this factorization, the "$SU(2)$ decomposition" due to each block structure ($U(1) \times SU(2)$ in reality). Then, Hilbert space $\mathcal{H}^{2^n}$ becomes the direct sum of $2^{n-1}$ subspaces generated for each associated pair $\{ \left| \alpha_{j(i)} \right>, \left| \alpha_{k(i)} \right> \}$, $i=1,2,...,2^{n-1}$. In each subspace, there is a complex mixing dynamics of probabilities, but without mix subspace probabilities.

\section{Bell gems basis fits naturally on $\left| \alpha_j \right>$}
For $n=2d$, Bell gems \cite{jaeg1,sych1} form a orthogonal basis of entangled states for $2d$ particles:

\begin{eqnarray}\label{gems}
\left| \Psi_{\mathcal{I}_4^{d}} \right> &=& \frac{1}{\sqrt{2^{d}}} \sum_{\{\epsilon_j\},\{\delta_k\}}(\tilde{\sigma}_{i_1} \otimes ... \otimes \tilde{\sigma}_{i_d})_{\epsilon_1 ... \epsilon_d,\delta_1 ... \delta_d} \left| \epsilon_1 ... \epsilon_d \right> \otimes \left| \delta_1 ... \delta_d \right> \\
&=& \frac{1}{\sqrt{2^{d}}} \sum_{\mathcal{E},\mathcal{D}=0}^{2^d-1}(\tilde{\sigma}_{i_1} \otimes ... \otimes \tilde{\sigma}_{i_d})_{\mathcal{E}_2^d,\mathcal{D}_2^d} \left| \mathcal{E}_2^d \right> \otimes \left| \mathcal{D}_2^d \right>
\end{eqnarray}

\noindent where $\{\epsilon_j\}=\{\epsilon_1,...,\epsilon_d\},\{\delta_k\}=\{\delta_1,...,\delta_d\}; \epsilon_j,\delta_k=0,1$. At this point, $\tilde{\sigma}_i$ can be considered as the traditional Pauli matrices. In addition, ${\mathcal{I}_4^{d}}$ is the set of digits of $\mathcal{I}$ when it is written in base-4 with $d$ digits ($\mathcal{I} \in \{0,1,...,4^d-1 \})$, it means $\{i_1,i_2,...,i_d\}$. In a similar way, $\mathcal{E}_2^d,\mathcal{D}_2^d$ are numbers written in base-2 with $d$ digits ($\mathcal{E},\mathcal{D} \in \{0,1,...,2^d-1 \})$. Then it is possible express the components of $H$ in that basis, obtaining a master expression to determine the necessary restrictions to fit their elements in the states $\{ \left| \alpha_{j(i)} \right>$ for the $SU(2)$ decomposition:

\begin{eqnarray}\label{hamiltalpha}
\left< \Psi_{\mathcal{I}_4^d} |H| \Psi_{\mathcal{K}_4^d} \right> &=& \frac{1}{2^{d}} \sum_{\mathcal{J}=0}^{4^{2d}-1} h_{\mathcal{J}_4^{2d}} \prod_{s=1}^d {\rm Tr} (\tilde{\sigma}_{i_s} \sigma_{j_{d+s}} \tilde{\sigma}^T_{k_s} \sigma^T_{j_s}) 
\end{eqnarray}

\noindent where $\mathcal{J} \in \{0,1,...,4^{2d}-1 \},\mathcal{I},\mathcal{K} \in \{0,1,...,4^{d}-1 \}$ (here, $\mathcal{J}=0$ can be removed in spite of the initial discussion). In last expressions, the product $\tilde{\sigma}_{i_s} \sigma_{j_{d+s}} \tilde{\sigma}^T_{k_s} \sigma^T_{j_s}$ has some properties in spite of Pauli matrices properties. Because $\sigma_1, \sigma_2, \sigma_3$ are traceless and $\sigma_i^T=\pm \sigma_i$ (negative sign only if $i=2$), then ${\rm Tr} (\tilde{\sigma}_{i_s} \sigma_{j_{d+s}} \tilde{\sigma}^T_{k_s} \sigma^T_{j_s})$ is non-zero only if: a) ${i_s}={j_{d+s}}={k_s}={j_s}$, b) ${i_s},{j_{d+s}},{k_s},{j_s}$ completely different between them, and c) ${i_s},{j_{d+s}},{k_s},{j_s}$ are equal by pairs.

A remark is convenient in this stage. In some works, as in \cite{sych1}, Bell gems are preferred be defined using $\tilde{\sigma}_i=\sigma_i$ for $i=0,1,3$ and $\tilde{\sigma}_2=i \sigma_2$ because it lets to have real coefficients when they are expressed in the computational basis $\left|0\right>, \left|1\right>$ (other alternative definitions can introduce specific phase factor in $\tilde{\sigma}_i$). We will adopt last definition in the following, which does not produce changes in the previous discussion. 

Last analysis conduces to a convenient definition: splitting the $2d$ set of particles, then two parts $i,j \in \{1,2,...,2d\}$, are {\it correspondents} if $j=i+d$ (they are in the same position of subscripts $1,2,...,d$ as the other in the second one $d+1,d+2,...,2d$). 
In last terms, a careful but direct analysis and development of (\ref{hamiltalpha}) shows that the $SU(2)$ decomposition arises when Hamiltonian depicts two types of interactions (by requiring ${\rm Tr} (\tilde{\sigma}_{i_s} \sigma_{j_{d+s}} \tilde{\sigma}^T_{k_s} \sigma^T_{j_s}) \ne 0$ for all $s$ in specific entries). The first one (Type I) comprehends all non local interactions between any correspondent parts in any direction. These terms generate diagonal entries in the Hamiltonian expressed in Bell gems basis. Together, it could be include two local interactions in one specific direction on only one pair up most of correspondent parts generating the off diagonal entries to conform the $SU(2)$ blocks. Note these local interaction terms could be interpreted as external driven fields as in \cite{delgadoA, delgadoB}. The second one (Type II) is obtained substituting the  Type I local interactions with non local interactions between pairs of any non correspondent parts included in exactly two pairs of correspondent parts. It means, if $k,k',k+d,k'+d$ with $k<k'\le d$ are these two pairs of correspondent parts, then interactions allowed are $(k,k'), (k,k'+d), (k',k+d)$ and $(k+d,k'+d)$. This group of four interactions generates the off diagonal entries to conform $SU(2)$ blocks. Type II interaction normally could be interpreted as a non driven process. Figure 1 resumes those two types of interactions.

\begin{figure}[h]
\begin{center}
\includegraphics[width=30pc]{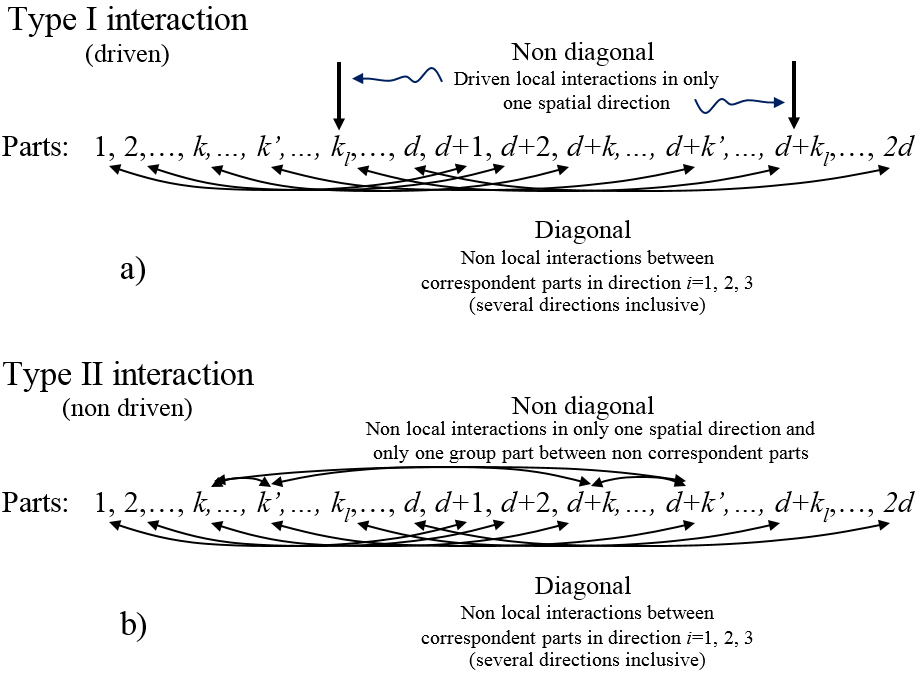}\hspace{1pc} \label{fig1}
\caption{Two types of physical interactions generating $SU(2)$ block decomposition: a) Type I interaction, and b) Type II interaction.}
\end{center}
\end{figure}

\section{Potential applications}
Some applications of $SU(2)$ decomposition are foreseen. It can be exploited in quantum control of bigger systems in which control schemes are not well developed as those of $SU(2)$ dynamics. The selectivity of pairing is related with the non diagonal elements arisen, it means, with the local interactions in Type I case and with no correspondent interactions in Type II case. This approach to quantum evolution will let control analytically the flow of quantum information in different geometrical arrangements. In a related but not necessarily equivalent direction, selective block decomposition could be useful for unitary factorization in quantum gate design \cite{delgadoC}. Finally, other applications in quantum superdense coding could be engineered for multichannel quantum information storage, using each subspace to storage differentiated information which could be necessary to process simultaneously, by example in quantum image processing.

\end{document}